\documentclass[fleqn,10pt]{wlscirep}
\usepackage{graphicx}
\usepackage{color}
\usepackage{hyperref}

\newcommand{\Tr}{\mathrm{Tr} \;}

\newcommand{\bk}{{\bf k}}

\renewcommand{\r}{\mathbf{r}}

\newcommand{\n}{\hat{n}}
\newcommand{\cd}{\hat{c}^\dagger}
\renewcommand{\c}{\hat{c}^{\phantom\dagger}}

\title{Quantum Thermalization and the Expansion of Atomic Clouds}
\author[1,*]{Louk Rademaker}
\author[2]{Jan Zaanen}
\affil[1]{Kavli Institute for Theoretical Physics, University of California Santa Barbara, CA 93106, USA}
\affil[2]{Institute-Lorentz for Theoretical Physics, Leiden University, P.O. Box 9506, Leiden, The Netherlands}

\affil[*]{louk.rademaker@gmail.com}

\begin{abstract}
The ultimate consequence of quantum many-body physics is that even the air we breathe is governed by strictly unitary time evolution. The reason that we perceive it nonetheless as a completely classical high temperature gas is due to the incapacity of our measurement machines to keep track of the dense many-body entanglement of the gas molecules. The question thus arises whether there are {\em instances where the quantum time evolution of a macroscopic system is qualitatively different from the equivalent classical system?}
Here we study this question through the expansion of noninteracting atomic clouds. While in many cases the full quantum dynamics is indeed indistinguishable from classical ballistic motion, we do find a notable exception. The subtle quantum correlations in a Bose gas approaching the condensation temperature appear to affect the expansion of the cloud, as if the system has turned into a diffusive collision-full classical system.
\end{abstract}
\begin{document}

\flushbottom
\maketitle
\thispagestyle{empty}

% introductory paragraph (not abstract) of approximately 150 words, summarizing the background, rationale, main results (introduced by "Here we show" or some equivalent phrase) and implications of the study. 

% current version: 146 words

\section*{Introduction}

\begin{figure}
  \includegraphics[width=0.7\textwidth]{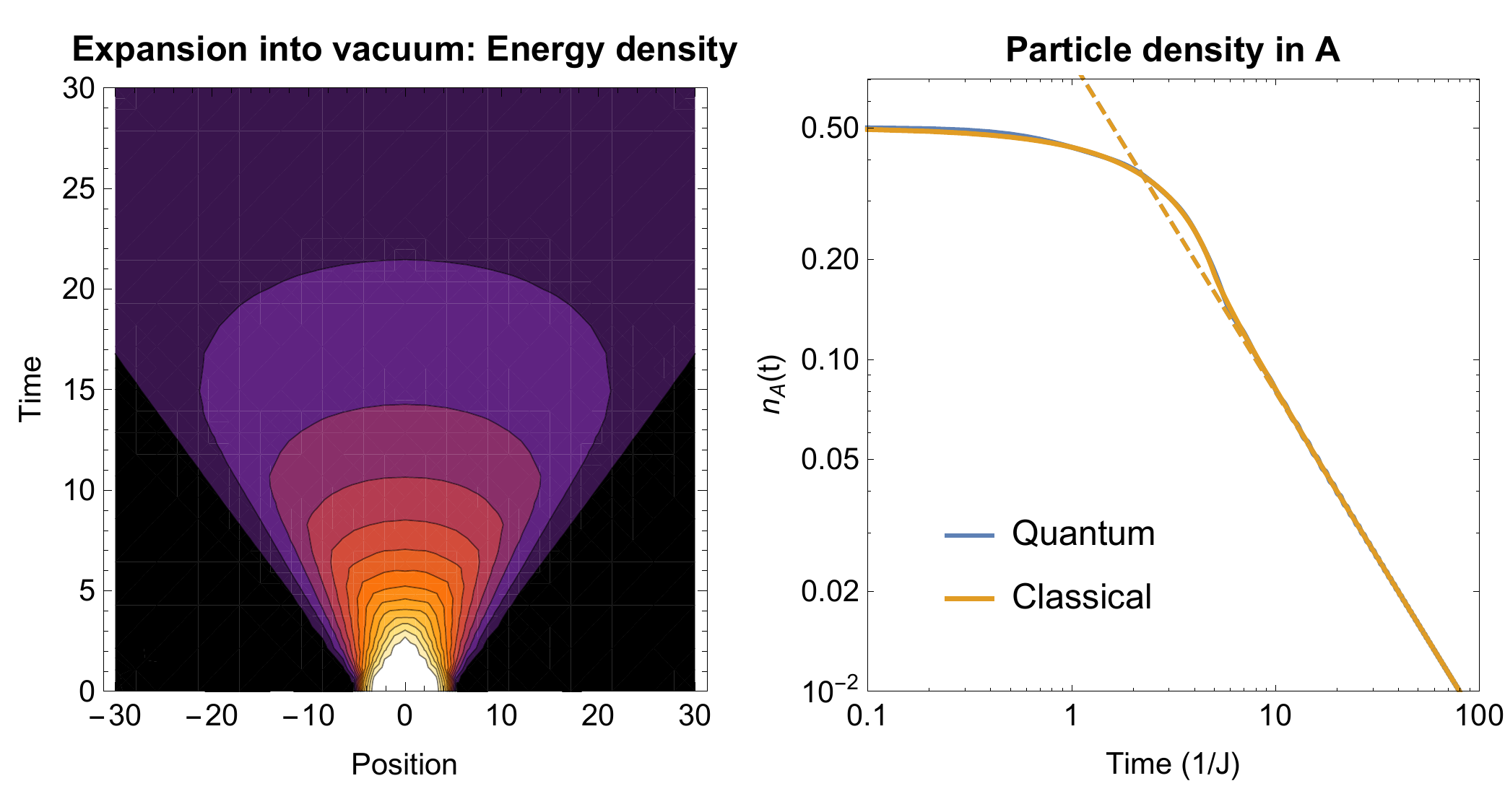}
  \caption{Expansion of an atomic cloud into the vacuum. Initially we prepare an one-dimensional atomic cloud in region $A$ ($|x|<5$), at inverse temperature $\beta = 0.01$, with particle density $n=0.5$. On the left we show the energy density as a function of position $x$ and time $t$, following exact quantum evolution. On the right we show the density of particles $n_A(t)$ in region $A$. The full quantum evolution is correctly represented by a classical distribution of particle positions and velocities $n(x,v,t)$ whose time evolution is given by ballistic motion, $n(x,v,t) = n(x-vt,v,0)$. The particle density in region $A$ decays as $n_A \sim 1/t$, in the right graph shown as a dashed line.}
  \label{FigVacuumExp}
\end{figure}

%%% main text intro

The laws describing classical gases, most notably the Second Law of Thermodynamics, seem at odds with the principle of unitary time evolution in quantum physics.\cite{clausius1856x} However, high energy states are densely many-body entangled and consequently the Eigenstate thermalization hypothesis (ETH)\cite{1991PhRvA..43.2046D,1994PhRvE..50..888S,Rigol:2011bf} claims that 
the outcomes of local measurements will be at long times indistinguishable from the outcome of the measurement in a thermal mixed state, at a temperature consistent with the energy of that state.\cite{2015CMaPh.340..499M,2015arXiv151203713D} 

Is this also true for a cloud of non-interacting quantum particles confined in a potential, which is suddenly released and allowed to expand in an infinite bath? This is actually similar to the key `time-of-flight measurement' in many cold atom experiments.\cite{Anderson:1995,1995PhRvL..75.3969D} After suddenly releasing the confining potential the atomic clouds 
expand, and by assuming that this is governed by ballistic, collision-less atomic motion the initial velocity distributions can be deduced from the expansion 
of the cloud. Invariably, it has been assumed that this expansion is governed by a purely classical Newtonian or wave kinematics, and this is undoubtedly a correct procedure to follow.

However, it is not at all obvious why this works. After all, before releasing the trapping potential, one may be in a quantum regime 
with Bose condensation or Fermi-degeneracy. How can these atoms suddenly behave like classical canon balls? In the next section, we will present a method to compute local observables exactly in the full quantum evolution by evaluating the logarithm of the density matrix. 
A first result is shown in Fig.\ref{FigVacuumExp}:  under the conditions of the cold atom experiments  the full quantum dynamics is indeed {\em indistinguishable} from classical ballistic expansion. 

\begin{figure*}
  \includegraphics[width=\textwidth]{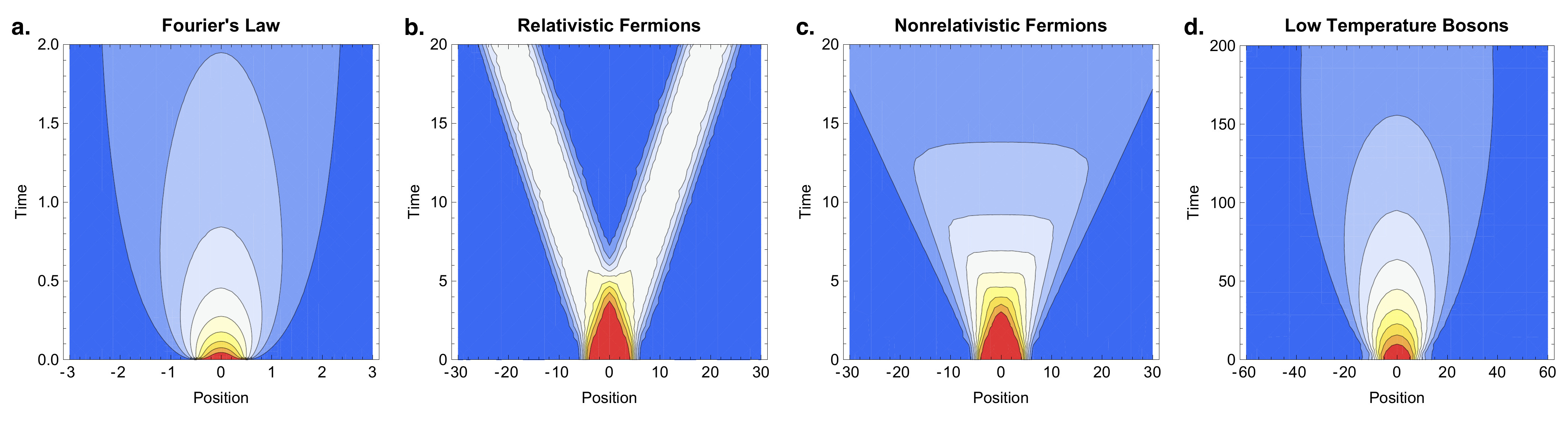}
  \caption{The energy spread of an initial subsystem $A$ at a hot temperature $T_A$ (in red) immersed in a cold bath at $T_B$ (in blue), for four different theories, computed in $d=1$. \textbf{a:} The classical Fourier's law predicts a diffusive spread of the heat, with the temperature difference in subsystem $A$ vanished according to $\Delta T(t) \sim t^{-1/2}$.
  \textbf{b:} For relativistic fermion systems ($T_A = 2, T_B = 1, n=0.5$) there is instantaneous thermalization once $A$ is in complete causal contact with the bath.
  \textbf{c:} In non-relativistic fermion systems ($T_A = 2, T_B = 1, n=0.5$) there is ballistic transport of particles, however, since not all particles have the same speed there is a power-law decay of the initial temperature difference, following $\Delta T(t) \sim t^{-1}$.
  \textbf{d:} Non-relativistic boson systems display a crossover from ballistic to diffusive thermalization. Here we show the energy profile in the low temperature regime where diffusive behavior is visible, $T_A = 100, T_B = 0.2$ and $n=0.5$.}
  \label{FigProfiles}
\end{figure*}

\begin{figure*}
  \includegraphics[width=\textwidth]{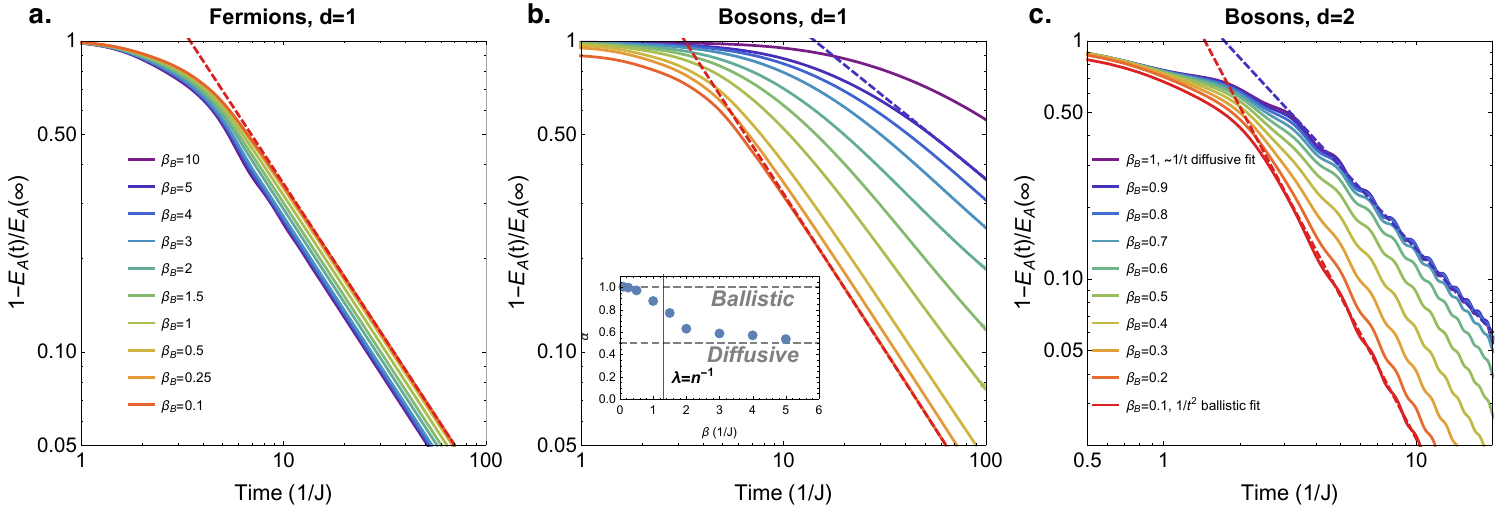}
  \caption{Decay of the energy difference between the system and the bath in non-relativistic fermionic or bosonic systems. In $d=1$ (left two pictures) we have immersed a subsystem $A$ at almost infinite temperature $\beta_A = 0.01$ in a bath with varying temperatures $\beta_B$. The chemical potential is tuned such that the particle density is $n=1/2$, and the total system size is $L=200$. On the vertical axis we plot the energy density in subsystem $A$, $E_A(t)$, normalized by the energy density at infinite time $E_A(t = \infty)$. In $d=2$ (right picture) the subsystem $A$ has size $N_A = 6 \times 6$ in a total system size of $N = 48 \times 48$.
  \textbf{a:} In fermionic systems, the decay is always of a ballistic nature, $\Delta E \sim t^{-d}$.
  \textbf{b:} In bosonic systems, there is a crossover from ballistic $t^{-d}$ to diffusive $t^{-d/2}$ decay. We fit the long-time behavior with the power-law form $t^{-\alpha}$. The inset shows the power $\alpha$ as a function of inverse bath temperature $\beta_B$. The crossover from ballistic to diffusive occurs around the point $\beta_c \sim 1.3$ where the thermal de Broglie wavelength $\lambda$ is comparable to the interparticle spacing $n^{-1}$, suggesting the wave-like nature of the bosons is responsible for the diffusive behavior.
  \textbf{c:} The crossover can also be observed in $d=2$ dimensions. The crossover occurs at higher temperatures, since the value where $\lambda \sim n^{-1/2}$ has shifted to higher temperature, $\beta_c \sim 0.8$. Small oscillations with period $1/4J$ can be observed due to the specific choice of lattice dispersion.}
  \label{FigEntEne}
\end{figure*}

We then address
cooling, where the atoms are released in a particle bath which is at a lower temperature than the trapped particles. 
When the temperature of the bath is high enough we find an expansion consistent with the classical expectation: since the particles do not collide, the hot cloud cools ballistically.
Similarly, when the cloud and the bath are both formed from fermions the system  behaves classical. 
However, for a cloud of bosons cooling into a bosonic bath at a temperature approaching the condensation temperature, the cooling is governed by {\em diffusion}! In Fig. \ref{FigProfiles} we show how the energy density of a 
 `hot' cloud in a cold bath spreads out in time, marking a clear difference between classical diffusion, ballistic fermionic behavior and again diffusion for a low temperature bosonic bath. 
 Quantitatively, the difference between ballistic and diffusive behavior can be shown by measuring the total energy density $\Delta E$ in the region of the original cloud relative to the bath energy density, 
 as shown in Fig. \ref{FigEntEne}: ballistic decay is characterized by $\Delta E \sim t^{-d}$ whereas diffusion satisfies $\Delta E \sim t^{-d/2}$.
 
This is our main result.  We have identified a circumstance where the quantum evolution becomes sharply distinguishable from the analogous classical evolution.  In the classical system diffusional expansion 
requires collisions,  but these are collision-less quantum particles.  We will explain how to test this prediction in cold atom experiments, but first we elucidate how these matters are computed.

%%%%%%%%%%%%%%%%%%%%%%%%%%%%%%%%%%%%%%%%%%%%%%%%%%%%%%%%%%%%%%%%
%% Here ends the introduction. 
%% Current length: 1002 words

\section*{Method}

The traditional approach to evaluate quantum time evolution is by repeated application of the time evolution operator $e^{-i \mathcal{H}dt}$ with small temporal steps $dt$. 
However, with this procedure it is impossible to reach times later than $t \sim 1/E$, where $E$ is a typical energy scale of the system. The hypothesis of thermalization provides us now with a simpler way to compute time evolution through the \emph{modular Hamiltonian} $\mathcal{M}$, which is the logarithm of the density matrix,
\begin{equation}
	\mathcal{M} = - \log \rho.
\end{equation}
As we will see, at late times $\mathcal{M}$ will simplify dramatically. Since we are interested in a hot cloud in a cold bath, our initial density matrix will have the form\footnote{This is equivalent, up to boundary terms, to $\rho_0 \sim \mathrm{Tr}_B\, e^{-\beta_A \mathcal{H}} \otimes \mathrm{Tr}_A\, e^{-\beta_B \mathcal{H}}$.}
\begin{equation}
	\rho_0 = \frac{1}{Z_AZ_B} e^{-\beta_A \mathcal{H}_A} \otimes e^{-\beta_B \mathcal{H}_B}
	\label{InitialState}
\end{equation}
where $\mathcal{H}_X$  and $\beta_X$  are the total Hamiltonian and inverse temperatures respectively, restricted to the  subsystems $X=A,B$. The time evolution of the modular Hamiltonian 
follows directly from the von Neumann equation for the time evolution of the density matrix,
\begin{equation}
	\mathcal{M}(t) =  e^{-i\mathcal{H}t} \mathcal{M}_0 e^{i\mathcal{H}t}.
\end{equation}
For noninteracting systems $\mathcal{H} = \sum_k \xi_k n_k$, the initial modular Hamiltonian following from Eqn. (\ref{InitialState}) can be written as, 
\begin{equation}
	\mathcal{M}_0 = \sum_{kk'} m_{kk'} \cd_k \c_{k'} + \log Z \;.
\end{equation}
The modular matrix $\hat{m} = m_{kk'}$ is Hermitian;  the sum runs over the momenta $k$ of the particles, while the constant $\log Z = - \eta \Tr \log \left(1 - \eta e^{-\hat{m}} \right)$, with $\eta = -1$ for fermions and
 $\eta = +1$ for bosons. The time evolution of both fermion and boson field operators appearing in the modular Hamiltonian is for the free system simply given by, 
\begin{equation}
	\cd_k (t) = e^{-i\mathcal{H}t} \cd_k e^{i \mathcal{H} t}
	 = e^{-i \xi_k t} \cd_k,
	\label{cd_timeevo}
\end{equation}
This implies for the time dependence of the modular Hamiltonian, 
\begin{eqnarray}
	\mathcal{M} (t) &=& \sum_{kk'} m_{kk'} e^{-i(\xi_k - \xi_{k'})t} \; \cd_k \c_{k'} + \log Z \\
		&\equiv& \sum_{kk'} m_{kk'} (t) \; \cd_k \c_{k'} + \log Z
\label{modHtimeev}		
\end{eqnarray}
It follows that time evolution corresponds with a unitary transformation on the modular matrix.  Local observables as the occupation numbers and the energy are in turn 
functions of the equal-time Greens function at time $t$, $G_{ij} (t) = \Tr \cd_i \c_j \rho(t)$, in terms of the modular matrix
\begin{equation}
	\hat{G} (t) = \left[ e^{\hat{m}(t)} - \eta \right]^{-1}.
	\label{GreensF}
\end{equation}
The advantage of this formulation starts to shimmer through. The intricacies of the 
full quantum evolution are absorbed in the strongly oscillating factors occurring in Eq. (\ref{modHtimeev}). These will rapidly average away such that in the limit $t \rightarrow \infty$, the modular Hamiltonian 
approaches the actual Hamiltonian, $\mathcal{M}(t) \rightarrow \beta \mathcal{H}$ when expressed in a {\em local basis}. 

\section*{Expansion of a noninteracting hot gas in a cold bath}

To see how this works let us consider some examples. Relativistic systems are discussed in the supplementary material\cite{suppl},  
reproducing the wisdom that these 
thermalize instantaneously once full causal contact is established\cite{Calabrese:2006bg,2016JSMTE..06.4003C,Bhaseen:2015et,2016PhRvD..94b5004L}. To model non-relativistic atoms
we resort to a lattice regularization in the form of  a hypercubic lattice in $d$ dimensions with nearest neighbor hopping,
\begin{equation}
	\mathcal{H} = - J \sum_{\r, \delta} \left( \cd_{\r+\delta} \c_\r + \cd_{\r} \c_{\r+\delta} \right)
		= \sum_\bk \epsilon_\bk \n_\bk
\end{equation}
with $\epsilon_\bk = -2 J \sum_{i=1}^d \cos k_i$. Given our initial hot cloud state the modular Hamiltonian equals $\mathcal{M}(t) = \beta_B \mathcal{H} + (\beta_A -\beta_B) \mathcal{H}_A(t)$ where the Hamiltonian of the subsystem $A$ is at $t=0$ equal to
\begin{equation}
	\mathcal{H}_A = - J \sum_{r_x, r_y, \ldots=1}^{L_A-1} \sum_\delta \left(  \cd_{\r+\delta} \c_\r + \cd_{\r} \c_{\r+\delta}  \right).
\end{equation}
Under time evolution this hot cloud spreads out and at $t>0$ we express $\mathcal{M}(t) = - J \sum_{j \ell} m_{j \ell}(t) \cd_j \c_\ell$ in terms of the elements of the modular matrix $m_{j \ell}(t)$ in the real space basis,
\begin{equation}
	m_{j \ell}(t)= \beta_B \delta_{|j-\ell| = 1}
	+ (\beta_A - \beta_B) 
	%\times
	%\nonumber \\ &&
	\int \frac{d^d \bk d^d \bk'}{(2\pi)^{2d}}
		e^{-i\bk \r_j + i \bk' \r_\ell} \,
		\left( \sum_{i=1}^d e^{i k_i} + e^{-ik'_i} \right) 
	%\nonumber \\ &&
		\left( \prod_{i=1}^d \frac{e^{i(k_i-k'_i)(L_A - 1)} - 1}{e^{i(k_i-k'_i)} - 1} \right)
			\, e^{i (\epsilon_\bk - \epsilon_{\bk'}) t}.
\end{equation}
Recall that thermalization in the ETH sense implies that the second term should vanish at late times. Indeed, using the continuum approximation $\epsilon_\bk \approx Jk^2 - \mu$ for $t \gg 1/\epsilon_\bk$, and thus $(\epsilon_\bk - \epsilon_{\bk'})t \approx (\bk + \bk')(\bk - \bk') J t$, we find for a site $j \in A$,
\begin{equation}
	\Delta m_{j, j+1} (t \gg 1) = 2 \Delta \beta (0)  \left( \frac{L_A -1}{2\pi Jt}\right)^d \sim \frac{V_A}{t^d}.
		\label{PowerLawThermoHighD}	
\end{equation}
Regardless of the statistics of the particles, the modular Hamiltonian approaches the final thermal state with a ballistic powerlaw decay.\cite{suppl}

However, the experimentally relevant {\em local energy density} in subsystem $A$ can approach the bath value in different manners, pending the quantum statistics of the particles as illustrated in Fig. \ref{FigEntEne}. 
Fermions are consistently subjected to a ballistic decay of the energy difference between the bath and the subsystem $A$ (Fig. 3a) and the resulting energy flow profile (Fig. 2c) displays a smoothened light-cone following 
the Lieb-Robinson bound with $v_{LR} = 2J$.\cite{1972CMaPh..28..251L} Turning to bosons, the surprise we announced becomes manifest:  we find a crossover from ballistic behavior at high bath temperatures to diffusive $\Delta E \sim t^{-d/2}$ at low bath temperatures. For both $d=1$ and $d=2$ dimensions (Fig. 3b,c), the crossover occurs around the point where the lattice thermal de Broglie wavelength corresponds to the interparticle spacing. This suggests that diffusive behavior is a consequence of the wave-like nature of the bosons, where $\partial_t \psi \sim \partial_x^2 \psi$. 

The energy profile of the diffusive case (Fig. 2d)  is surprisingly reminiscent of the classical Fourier's law of heat diffusion (Fig. 2a). However, one should not be fooled by this apparent relation to classical diffusion. After all, we are considering noninteracting particles and the equivalent classical description of our set-up is through a distribution of particles and velocities $n(x,v,t)$ that evolves ballistically $n(x,v,t) = n(x-vt,v,0)$. For the expansion into a cold bosonic bath the classical picture still yields a ballistic spread, while the exact quantum evolution displays diffusive behavior.\cite{suppl} The diffusive behavior for cold bosonic baths is therefore a genuine quantum effect.

\begin{figure}
  \includegraphics[width=0.5\textwidth]{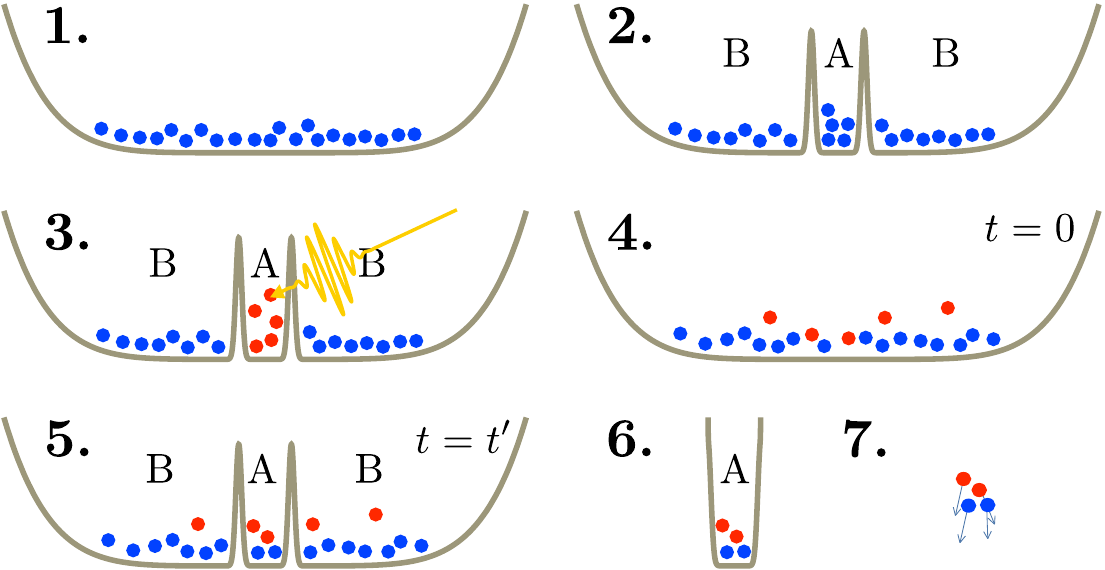}
  \caption{Cartoon of the suggested cold-atom experiment. 1. Prepare a trapped cloud of noninteracting atoms at a temperature $T_B$. 2. Introduce a barrier that separates the system into $A$ and a bath $B$. 3. Using a laser, heat up the atoms in region $A$ to temperature $T_A$. 4. At time $t=0$ remove the barrier between $A$ and $B$, letting the two subsystems thermalize. 5. After a time $t=t'$ reintroduce the barrier between $A$ and $B$. 6. Remove the trap around region $B$. 7. Remove the trap around $A$ and perform a time-of-flight measurement of the kinetic energy of the atoms in $A$. The steps 1-7 should be repeated for different times $t'$ to obtain the energy in $A$ as a function of time after the quench.}
  \label{FigExperiment}
\end{figure}

\section*{Conclusion and Outlook}

This interesting crossover can be probed directly in experiments using cold atoms, following the protocol illustrated in Fig.~\ref{FigExperiment}.\cite{Bloch:2008gl,Polkovnikov:2016dj,Jean:2016jm} Initially, one prepares a cloud of atoms tuned to be noninteracting using the Feshbach resonance. Using optical lattice techniques a barrier is created in between $A$ and $B$, and a separate laser excites $A$ to be at a different temperature than $B$. At time $t=0$ the barrier is removed and the system will evolve as described. To measure the energy density in subsystem $A$ after a time $t$, one reintroduces the barrier, let the atoms in the bath $B$ escape, followed by  time-of-flight measurements of the distribution of momenta of the atoms in $A$. From the distribution of these momenta the total kinetic energy can be reconstructed. The experiment is then repeated to obtain the energy density in $A$ at every time instance. In this way the curves of Fig. \ref{FigEntEne}, for either ballistic or diffusive behavior, can be experimentally measured.

It might be a surprise to observe thermalization in integrable non-interacting systems, but it is quite straightforward that this happens for local quenches such as the one studied here.\cite{Eisert:2015ka,Cramer:2008ca} Even though there are many integrals of motion, there is no conservation law that restricts certain degrees of freedom to remain within $A$. Note, however, that systems where the integrals of motion are truly local, as is the case for Anderson insulators\cite{Anderson:1958fz} or the many-body localized phase\cite{Huse:2014co,Nandkishore:2015kt}, information remains within $A$ and no thermalization will occur.

A critical reader might object that the system we study actually displays an entropy decrease. However, much like refrigerators, we reduce the entropy of subsystem $A$ by increasing the bath entropy by at least the same amount. In fact, while the total entropy remains constant in any quantum system, the mutual information $\mathcal{I}_{AB} (t)= S_A (t)+ S_B (t)- S_{A \cup B}$ increases upon thermalization since the subsystem $A$ and the bath $B$ become entangled. This increase in mutual information should be considered the quantum version of the Second Law.\cite{clausius1856x} However, it remains an open question to prove this increase for thermodynamically large systems as the Second Law requires.

%%%%%%

\section*{Acknowledgements}

We are thankful to Tarun~Grover, Laimei~Nie, Mike~Zaletel and Immanuel~Bloch for discussions. L.R. was supported by the Dutch Science Foundation (NWO) through a Rubicon grant and by the National Science Foundation under Grant No. PHY11-25915 and Grant No. NSF-KITP-17-019.

\section*{Author contributions statement}

L.R. performed the numerics, and L.R. and J.Z. wrote the manuscript together.

\section*{Additional information}
The authors declare no competing financial interests.

\newpage

\appendix

\renewcommand\thefigure{\thesection.\arabic{figure}}   
%%%%%%%%%%%%%%%%%%%%%%%%%%%%%%%%%%%%%%%%%%%%%%%%%%%%%%%%%%%%
\section{Thermalization of classical systems - Fourier's Law}
\label{SecFourier}

In the main manuscript we consider a hot system $A$ at temperature $T_A$ immersed in a cold bath at temperature $T_B$. How will thermal equilibrium be reached according to the classical theory of thermal diffusion? There it is assumed that a system is locally in thermal equilibrium, such that one can define a temperature $T(x)$ at each point in space. If there exists a temperature gradient, energy will flow from hot to cold according to Fourier's law,
\begin{equation}
	j_E (x) = - \kappa(T(x)) \; \nabla T(x)
\end{equation}
where $\kappa$ is the thermal conductivity of the material and $j_E(x)$ is the energy current. If we are in a regime where both the specific heat $c_V$ and the thermal conductivity $\kappa$ are approximately independent of temperature, Fourier's Law becomes a diffusion equation
\begin{equation}
	\partial_t T =  \mathcal{D} \nabla^2 T.
\end{equation}
where the diffusion constant is $\mathcal{D} =\frac{\kappa}{c_V}$. This equation can be solved using the heat kernel.

Let us look explicitly at an initial state with a hot cloud at temperature $T_A$ for $|x| < a/2$, and and a bath at $T_B$ for $|x|>a/2$. The resulting solution of the heat diffusion equation yields
\begin{equation}
	\Delta T(x,t) = \frac{1}{2} (T_A - T_B) \left( \mathrm{Erf} \left[ \frac{a-2x}{4 \sqrt{\mathcal{D}t}} \right]
		+\mathrm{Erf} \left[ \frac{a+2x}{4 \sqrt{\mathcal{D}t}} \right] \right).
	\label{HydroSol}
\end{equation}
In the left panel of Fig. 1 in the main manuscript, we show how the heat of our cloud spreads according to the above formula.

Note that the temperature difference at long times falls of in a power law fashion, $\Delta T(x=0,t\gg1) \sim  \frac{a(T_A -T_B)}{2 \sqrt{\pi \mathcal{D} t}}$. In higher dimensions $d$, the above equations straightforwardly generalize to
\begin{equation}
	\Delta T(x=0,t\gg1) \sim  (T_A - T_B) \left( \frac{a}{2 \sqrt{\pi \mathcal{D} t}}\right)^{d} \sim \frac{V_A}{t^{d/2}}.
	\label{FLaw}
\end{equation}
Therefore, if the energy of temperature of a system decays as a powerlaw $t^{-d/2}$, we call this diffusion.

%%%%%%%%%%%%%%%%%%%%%%%%%%%%%%%%%%%%%%%%%%%%%%%%%%%%%%%%%%%%
\section{Comparison between classical and quantum description}

Now consider another classical system: a gas of collision-less non-interacting particles. At time $t=0$, we can characterize this gas as having a distribution of particles in position and velocity, $n(x,v,0)$. The particle density as a function of position is $n(x) = \int dv \; n(x,v,0)$, and with an energy per particle that depends only on velocity, $\epsilon(v)$, the energy density is given by $E(x) = \int dv \; \epsilon(v) n(x,v,0)$. 

Because the particles are collision-less and have no further interactions, the velocity is conserved. This means that the full distribution at a later time $t$ can be expressed in terms of the initial distribution as
\begin{equation}
	n(x,v,t) = n(x-vt,v,0).
\end{equation}
To model a generic system of bosons as is done in the main manuscript, we can start with an initial distribution
\begin{equation}
	n(x,v,0) = \frac{1}{2\pi} \frac{dk}{dv} \frac{1}{e^{\beta(x) (\epsilon(v) - \mu(x))} - 1 }
	\label{ClassBallP}
\end{equation}
with $\epsilon_\bk$ the boson dispersion and $v_\bk = \frac{d \epsilon_\bk}{d\bk}$. Note that the energy per particle is $\epsilon_\bk - \mu$. The initial temperature imbalance is characterized by a spatially varying inverse temperature $\beta(x)$ and chemical potential $\mu(x)$.

In the main manuscript we first considered a gas expanding into the vacuum. We model this in $d=1$ by taking $n(x,v,0) = \frac{m}{2\pi} \frac{1}{e^{\beta (mv^2/2 - \mu)}-1}$ when $x \in A$, and zero outside $A$. For bosons $\mu < 0$, so let's define $\alpha = e^{-\beta \mu}>1$. The particle density at $x=0$ at late time $t \gg 1$ then equals
\begin{eqnarray}
	n(x=0,t) &=& 2 \int_0^{L_A/2t} dv \frac{m}{2 \pi}  \frac{1}{e^{\beta (mv^2/2 - \mu)}-1} \\
		&\approx& \frac{m}{\pi} \int_0^{L_A/2t} dv \frac{1}{(\alpha - 1) + \beta m v^2/2} \\
		&\approx& \frac{mL_A}{2\pi (\alpha-1)} t^{-1} + \mathcal{O}(t^{-3}).
\end{eqnarray}
Similarly, the energy of bosons $\epsilon_\bk - \mu$ is always positive and nonzero which implies that the late-time behavior of the energy is $E(x=0,t) = \int dv (\epsilon(v) - \mu) n(-vt,0) \sim t^{-1}$. Therefore, whenever a system thermalizes with a powerlaw $t^{-d}$, we will call this ballistic behavior.

\begin{figure}
  \includegraphics[width=0.7\textwidth]{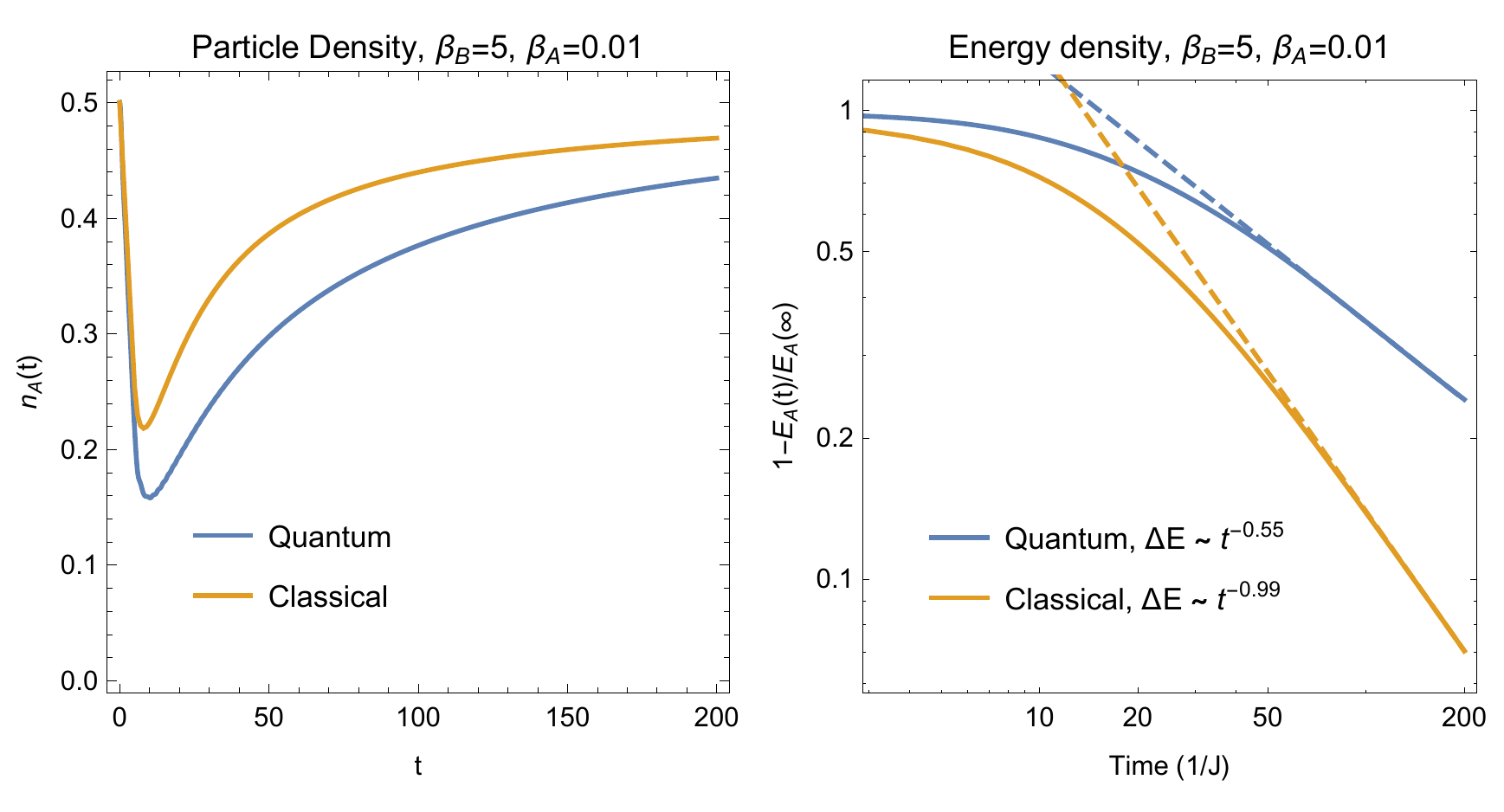}
  \caption{Comparison between quantum and classical description of the expansion of a hot $\beta_A = 0.01$ boson system into a cold bath $\beta_B=5$, with $L_A = 10$ and $n=0.5$. The quantum description is slower and tends to diffusive behavior $\Delta E \sim t^{-1/2}$, whereas the classical description incorrectly predicts ballistic behavior. }
  \label{FigClassQCompare}
\end{figure}

Finally, in the main manuscript we show that the correct quantum description of a hot bosonic system $A$ in a cold bosonic bath displays {\em diffusive} rather than ballistic behavior. In Fig. \ref{FigClassQCompare} we compare the results of this quantum thermalization to the classical ballistic picture following Eqn. (\ref{ClassBallP}) at $\beta_B = 5$. The classical picture incorrectly yields a ballistic spread, while the exact quantum computation displays diffusive behavior. The diffusive behavior for cold bosonic baths is therefore a true quantum effect.

\section{Entropy}

Entropy plays an important role in quantum many-body physics. Fortunately, the total entropy of the system, which is of course time-independent, is relatively easy to compute using the modular Hamiltonian,
\begin{equation}
	S = - \Tr \rho(t) \log \rho(t) = \Tr \mathcal{M}(t) \rho(t).
\end{equation}
For a free system, this implies that the entropy can be expressed in term of the eigenvalues of the Greens function $g_\alpha$ as 
\begin{equation}
	S = \sum_\alpha \left[ - g_\alpha \log g_\alpha + \eta (1+ \eta g_\alpha) \log (1 + \eta g_\alpha) \right],
	\label{EntropyEqnGF}
\end{equation}
where $\eta=\pm1$ is the sign for bosons/fermions. 

To obtain the entanglement entropy of the subsystem $A$, we need the reduced density matrix $\rho_A(t) = \Tr_{B} \rho(t)$. However, for free systems the entanglement entropy can be computed simply by using Eqn. (\ref{EntropyEqnGF}) where $g_\alpha$ are now the eigenvalues of the Greens function restricted to subsystem $A$.

\section{Thermalization of relativistic fermions}

There is an extensive literature on thermal quenches and thermalization of relativistic particles.\cite{Calabrese:2006bg,2016arXiv160302889C,Bhaseen:2015et,Lucas:2015uk} In one dimension, the Hamiltonian for relativistic fermions is
\begin{equation}
	\mathcal{H}
	= \int dx \left( \psi^\dagger_R(x) i v \partial_x \psi_R(x) - \psi^\dagger_L(x) i v \partial_x \psi_L(x) \right)
\end{equation}
where $\psi_{L,R}(x)$ is the field for left- and right-moving particles, respectively, and $v>0$ is the Fermi velocity. The energy of the right-movers is $\epsilon^R_k = v k$ and of left-movers is $\epsilon^L_k = -vk$. The right-moving nature of the $\psi_R$ field becomes obvious when one expresses the time evolution of the operator,
\begin{equation}
	\psi^\dagger_R(x,t) = \int \frac{dk}{2\pi} \psi^\dagger_{R,k} e^{ik(x-vt)}.
	\label{TimeEvoContRM}
\end{equation}
The initial modular Hamiltonian for our hot cloud in $A$ immersed in a bath reads
\begin{equation}
	\mathcal{M}_0 = \beta_B \mathcal{H} + (\beta_A - \beta_B) \int_{0}^{L_A} dx \, h(x) 
\end{equation}
Consider only the right-moving particles $h_R(x,t) = \psi^\dagger_R(x,t) i v \partial_x \psi_R(x,t)$ in subsystem $A$. Under unitary time evolution this segment shifts in its enterity to the right,
\begin{equation}
	\int_0^{L_A} dx \; h_R(x,t)
	=\int_{vt}^{L_A+vt} dx \;\psi^\dagger_R(x) i v \partial_x \psi_R(x)
\end{equation}
Similarly, the left-movers move to the left under time evolution. This means that after a time $t = L_A/v$ there are no remnants of the hot cloud left in the subsystem $A$. The moment the system $A$ has reached a full causal contact with the bath, it is {\em instantaneously thermalized}. In Fig. 1 of the main manuscript, middle left, we display the heat profile of such a relativistic system. Notice also that this procedure correctly reproduces the non-equilibrium steady state (NESS) $\mathcal{M} = \beta_+ H + \beta_- P$ as described in Ref. \cite{Bhaseen:2015et}.

\section{Decay of modular matrix for nonrelativistic particles}

\begin{figure}[t]
  \includegraphics[width=0.5\textwidth]{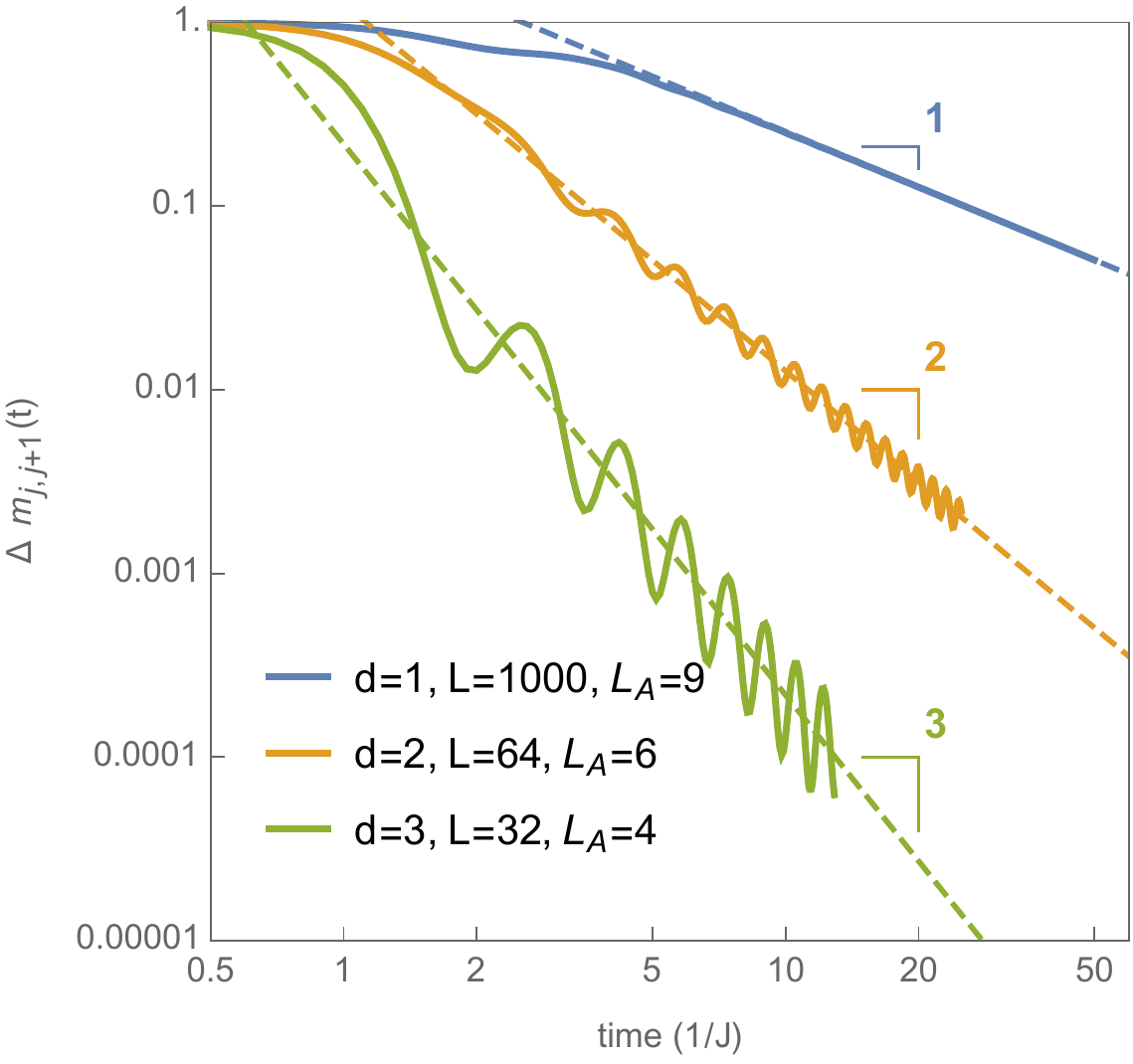}
  \caption{The decay of the nearest-neighbor modular matrix element $\Delta m_{j,j+1}(t)$ for $j \in A$, for $d=1,2$ and 3. The thick solid lines indicate the numerical results, solving Eqn. (11) of the main manuscript exactly on finite size systems. All approach the general formula describing ballistic decay Eqn. (12), here shown as dashed lines.}
  \label{FigBetaDecayAny}
\end{figure}

In the main manuscript we showed using the continuum approximation $\epsilon_\bk \approx Jk^2 - \mu$ that the matrix elements of the modular matrix decay ballistically, $\Delta m \sim t^{-d}$. In Fig. \ref{FigBetaDecayAny} we indeed show that this is the case when computed numerically on large finite size systems. The solid lines are the exact results, and the dashed line is the continuum limit. Note that there are oscillations visible with angular frequency $4J$, which is the result of corrections due to the lattice dispersion $\epsilon_\bk = -2 J \sum_{i=1}^d \cos k_i$. The precise shape and amplitude of these oscillations depend on the precise form of the dispersion.


\begin{thebibliography}{10}
\expandafter\ifx\csname url\endcsname\relax
  \def\url#1{\texttt{#1}}\fi
\expandafter\ifx\csname urlprefix\endcsname\relax\def\urlprefix{URL }\fi
\providecommand{\bibinfo}[2]{#2}
\providecommand{\eprint}[2][]{\url{#2}}

\bibitem{clausius1856x}
\bibinfo{author}{Clausius, R.}
\newblock \bibinfo{title}{{X. On a modified form of the second fundamental
  theorem in the mechanical theory of heat}}.
\newblock \emph{\bibinfo{journal}{The London, Edinburgh, and Dublin
  Philosophical Magazine and Journal of Science}}
  \textbf{\bibinfo{volume}{12}}, \bibinfo{pages}{81}
  (\bibinfo{year}{1856}).

\bibitem{1991PhRvA..43.2046D}
\bibinfo{author}{Deutsch, J.~M.}
\newblock \bibinfo{title}{{Quantum statistical mechanics in a closed system}}.
\newblock \emph{\bibinfo{journal}{Phys. Rev. A}} \textbf{\bibinfo{volume}{43}},
  \bibinfo{pages}{2046} (\bibinfo{year}{1991}).

\bibitem{1994PhRvE..50..888S}
\bibinfo{author}{Srednicki, M.}
\newblock \bibinfo{title}{{Chaos and quantum thermalization}}.
\newblock \emph{\bibinfo{journal}{Phys. Rev. E}} \textbf{\bibinfo{volume}{50}},
  \bibinfo{pages}{888} (\bibinfo{year}{1994}).

\bibitem{Rigol:2011bf}
\bibinfo{author}{Rigol, M.}, \bibinfo{author}{Dunjko, V.} \&
  \bibinfo{author}{Olshanii, M.}
\newblock \bibinfo{title}{{Thermalization and its mechanism for generic
  isolated quantum systems}}.
\newblock \emph{\bibinfo{journal}{Nature}} \textbf{\bibinfo{volume}{452}},
  \bibinfo{pages}{854} (\bibinfo{year}{2008}).

\bibitem{2015CMaPh.340..499M}
\bibinfo{author}{M{\"u}ller, M.~P.}, \bibinfo{author}{Adlam, E.},
  \bibinfo{author}{Masanes, L.} \& \bibinfo{author}{Wiebe, N.}
\newblock \bibinfo{title}{{Thermalization and Canonical Typicality in
  Translation-Invariant Quantum Lattice Systems}}.
\newblock \emph{\bibinfo{journal}{Communications in Mathematical Physics}}
  \textbf{\bibinfo{volume}{340}}, \bibinfo{pages}{499}
  (\bibinfo{year}{2015}).

\bibitem{2015arXiv151203713D}
\bibinfo{author}{Doyon, B.}
\newblock \bibinfo{title}{{Thermalization and pseudolocality in extended
  quantum systems}}.
 \bibinfo{pages}{arXiv:1512.03713}
  (\bibinfo{year}{2015}).

\bibitem{Anderson:1995}
\bibinfo{author}{Anderson, M.~H.}, \bibinfo{author}{Ensher, J.~R.},
  \bibinfo{author}{Matthews, M.~R.}, \bibinfo{author}{Wieman, C.~E.} \&
  \bibinfo{author}{Cornell, E.~A.}
\newblock \bibinfo{title}{{Observation of Bose-Einstein Condensation in a
  Dilute Atomic Vapor}}.
\newblock \emph{\bibinfo{journal}{Science}} \textbf{\bibinfo{volume}{269}},
  \bibinfo{pages}{198} (\bibinfo{year}{1995}).

\bibitem{1995PhRvL..75.3969D}
\bibinfo{author}{Davis, K.~B.} \emph{et~al.}
\newblock \bibinfo{title}{{Bose-Einstein Condensation in a Gas of Sodium
  Atoms}}.
\newblock \emph{\bibinfo{journal}{Phys. Rev. Lett.}}
  \textbf{\bibinfo{volume}{75}}, \bibinfo{pages}{3969}
  (\bibinfo{year}{1995}).

\bibitem{suppl}
See online supplementary information.

\bibitem{Calabrese:2006bg}
\bibinfo{author}{Calabrese, P.} \& \bibinfo{author}{Cardy, J.}
\newblock \bibinfo{title}{{Time Dependence of Correlation Functions Following a
  Quantum Quench}}.
\newblock \emph{\bibinfo{journal}{Phys. Rev. Lett.}}
  \textbf{\bibinfo{volume}{96}}, \bibinfo{pages}{136801}
  (\bibinfo{year}{2006}).

\bibitem{2016JSMTE..06.4003C}
\bibinfo{author}{Calabrese, P.} \& \bibinfo{author}{Cardy, J.}
\newblock \bibinfo{title}{{Quantum quenches in 1+1 dimensional conformal field theories}}.
\newblock \emph{\bibinfo{journal}{J. Stat. Mech.}}
  \textbf{\bibinfo{volume}{06}}, \bibinfo{pages}{064003}
  (\bibinfo{year}{2016}).

\bibitem{Bhaseen:2015et}
\bibinfo{author}{Bhaseen, M.~J.}, \bibinfo{author}{Doyon, B.},
  \bibinfo{author}{Lucas, A.} \& \bibinfo{author}{Schalm, K.}
\newblock \bibinfo{title}{{Energy flow in quantum critical systems far
  from~equilibrium}}.
\newblock \emph{\bibinfo{journal}{Nat. Phys.}} \textbf{\bibinfo{volume}{11}},
  \bibinfo{pages}{509} (\bibinfo{year}{2015}).

\bibitem{2016PhRvD..94b5004L}
\bibinfo{author}{Lucas, A.}, \bibinfo{author}{Schalm, K.},
  \bibinfo{author}{Doyon, B.} \& \bibinfo{author}{Bhaseen, M.~J.}
\newblock \bibinfo{title}{{Shock waves, rarefaction waves, and nonequilibrium
  steady states in quantum critical systems}}.
\newblock \emph{\bibinfo{journal}{Phys. Rev. D}} \textbf{\bibinfo{volume}{94}},
  \bibinfo{pages}{025004} (\bibinfo{year}{2016}).

\bibitem{1972CMaPh..28..251L}
\bibinfo{author}{Lieb, E.~H.} \& \bibinfo{author}{Robinson, D.~W.}
\newblock \bibinfo{title}{{The finite group velocity of quantum spin systems}}.
\newblock \emph{\bibinfo{journal}{Communications in Mathematical Physics}}
  \textbf{\bibinfo{volume}{28}}, \bibinfo{pages}{251}
  (\bibinfo{year}{1972}).

\bibitem{Bloch:2008gl}
\bibinfo{author}{Bloch, I.}, \bibinfo{author}{Dalibard, J.} \&
  \bibinfo{author}{Zwerger, W.}
\newblock \bibinfo{title}{{Many-body physics with ultracold gases}}.
\newblock \emph{\bibinfo{journal}{Rev. Mod. Phys.}}
  \textbf{\bibinfo{volume}{80}}, \bibinfo{pages}{885}
  (\bibinfo{year}{2008}).

\bibitem{Polkovnikov:2016dj}
\bibinfo{author}{Polkovnikov, A.} \& \bibinfo{author}{Sels, D.}
\newblock \bibinfo{title}{{Thermalization in small quantum systems}}.
\newblock \emph{\bibinfo{journal}{Science}} \textbf{\bibinfo{volume}{353}},
  \bibinfo{pages}{752} (\bibinfo{year}{2016}).

\bibitem{Jean:2016jm}
\bibinfo{author}{Kaufman, A.~M.} \emph{et~al.}
\newblock \bibinfo{title}{{Quantum thermalization through entanglement in an
  isolated many-body system}}.
\newblock \emph{\bibinfo{journal}{Science}} \textbf{\bibinfo{volume}{353}},
  \bibinfo{pages}{794} (\bibinfo{year}{2016}).

\bibitem{Eisert:2015ka}
\bibinfo{author}{Eisert, J.}, \bibinfo{author}{Friesdorf, M.} \&
  \bibinfo{author}{Gogolin, C.}
\newblock \bibinfo{title}{{Quantum many-body systems out of equilibrium}}.
\newblock \emph{\bibinfo{journal}{Nat. Phys.}} \textbf{\bibinfo{volume}{11}},
  \bibinfo{pages}{124} (\bibinfo{year}{2015}).

\bibitem{Cramer:2008ca}
\bibinfo{author}{Cramer, M.}, \bibinfo{author}{Dawson, C.~M.},
  \bibinfo{author}{Eisert, J.} \& \bibinfo{author}{Osborne, T.~J.}
\newblock \bibinfo{title}{{Exact Relaxation in a Class of Nonequilibrium
  Quantum Lattice Systems}}.
\newblock \emph{\bibinfo{journal}{Phys. Rev. Lett.}}
  \textbf{\bibinfo{volume}{100}}, \bibinfo{pages}{030602}
  (\bibinfo{year}{2008}).

\bibitem{Anderson:1958fz}
\bibinfo{author}{Anderson, P.~W.}
\newblock \bibinfo{title}{{Absence of Diffusion in Certain Random Lattices}}.
\newblock \emph{\bibinfo{journal}{Phys. Rev.}} \textbf{\bibinfo{volume}{109}},
  \bibinfo{pages}{1492} (\bibinfo{year}{1958}).

\bibitem{Huse:2014co}
\bibinfo{author}{Huse, D.~A.}, \bibinfo{author}{Nandkishore, R.} \&
  \bibinfo{author}{Oganesyan, V.}
\newblock \bibinfo{title}{{Phenomenology of fully many-body-localized
  systems}}.
\newblock \emph{\bibinfo{journal}{Phys. Rev. B}} \textbf{\bibinfo{volume}{90}},
  \bibinfo{pages}{174202} (\bibinfo{year}{2014}).

\bibitem{Nandkishore:2015kt}
\bibinfo{author}{Nandkishore, R.} \& \bibinfo{author}{Huse, D.~A.}
\newblock \bibinfo{title}{{Many-Body Localization and Thermalization in Quantum
  Statistical Mechanics}}.
\newblock \emph{\bibinfo{journal}{Annu. Rev. Condens. Matter Phys.}}
  \textbf{\bibinfo{volume}{6}}, \bibinfo{pages}{15} (\bibinfo{year}{2015}).

\end{thebibliography}

\begin{thebibliography}{10}
\expandafter\ifx\csname url\endcsname\relax
  \def\url#1{\texttt{#1}}\fi
\expandafter\ifx\csname urlprefix\endcsname\relax\def\urlprefix{URL }\fi
\providecommand{\bibinfo}[2]{#2}
\providecommand{\eprint}[2][]{\url{#2}}


\bibitem{Calabrese:2006bg}
\bibinfo{author}{Calabrese, P.} \& \bibinfo{author}{Cardy, J.}
\newblock \bibinfo{title}{{Time Dependence of Correlation Functions Following a
  Quantum Quench}}.
\newblock \emph{\bibinfo{journal}{Phys. Rev. Lett.}}
  \textbf{\bibinfo{volume}{96}}, \bibinfo{pages}{136801}
  (\bibinfo{year}{2006}).

\bibitem{2016JSMTE..06.4003C}
\bibinfo{author}{Calabrese, P.} \& \bibinfo{author}{Cardy, J.}
\newblock \bibinfo{title}{{Quantum quenches in 1+1 dimensional conformal field theories}}.
\newblock \emph{\bibinfo{journal}{J. Stat. Mech.}}
  \textbf{\bibinfo{volume}{06}}, \bibinfo{pages}{064003}
  (\bibinfo{year}{2016}).

\bibitem{Bhaseen:2015et}
\bibinfo{author}{Bhaseen, M.~J.}, \bibinfo{author}{Doyon, B.},
  \bibinfo{author}{Lucas, A.} \& \bibinfo{author}{Schalm, K.}
\newblock \bibinfo{title}{{Energy flow in quantum critical systems far
  from~equilibrium}}.
\newblock \emph{\bibinfo{journal}{Nat. Phys.}} \textbf{\bibinfo{volume}{11}},
  \bibinfo{pages}{509} (\bibinfo{year}{2015}).

\bibitem{2016PhRvD..94b5004L}
\bibinfo{author}{Lucas, A.}, \bibinfo{author}{Schalm, K.},
  \bibinfo{author}{Doyon, B.} \& \bibinfo{author}{Bhaseen, M.~J.}
\newblock \bibinfo{title}{{Shock waves, rarefaction waves, and nonequilibrium
  steady states in quantum critical systems}}.
\newblock \emph{\bibinfo{journal}{Phys. Rev. D}} \textbf{\bibinfo{volume}{94}},
  \bibinfo{pages}{025004} (\bibinfo{year}{2016}).
\end{thebibliography}
\end{document}